\documentclass[aps,pra,superscriptaddress,preprint]{revtex4-1}

\usepackage{graphicx}
\usepackage{color}
\usepackage{amsmath,amssymb}
\usepackage{bm}
\usepackage{upgreek}
\usepackage{xspace}
\usepackage{natbib}
\usepackage[colorlinks,urlcolor=blue,citecolor=blue,linkcolor=blue]{hyperref}

\newcommand{\vect}[1]{\mathbf{#1}\xspace}
\newcommand{\uvect}[1]{\hat{\mathbf{#1}}\xspace}

\begin{document}

\title{Magnetic pseudo-fields in a rotating electron-nuclear spin system}

\author{A.~A.~Wood}
\affiliation{School of Physics, University of Melbourne, Parkville, VIC 3010, Australia.}
\author{E. Lilette}
\affiliation{School of Physics, University of Melbourne, Parkville, VIC 3010, Australia.}
\author{Y. Y. Fein}
\affiliation{School of Physics, University of Melbourne, Parkville, VIC 3010, Australia.}
\author{V. S. Perunicic}
\affiliation{School of Physics, University of Melbourne, Parkville, VIC 3010, Australia.}
\author{{L.~C.~L.~Hollenberg}}
\affiliation{School of Physics, University of Melbourne, Parkville, VIC 3010, Australia.}
\affiliation{Center for Quantum Computation and Communication Technology, University of Melbourne, Parkville, Victoria 3010, Australia.}
\author{R. E. Scholten}
\affiliation{School of Physics, University of Melbourne, Parkville, VIC 3010, Australia.}
\author{A. M. Martin}
\email{martinam@unimelb.edu.au}
\affiliation{School of Physics, University of Melbourne, Parkville, VIC 3010, Australia.}

\date{\today}
\begin{abstract}
{\bf{A precessing spin observed in a rotating frame of reference appears frequency-shifted, an effect analogous to the precession of a Foucault pendulum observed on the rotating Earth. This frequency shift can be understood as arising from a magnetic pseudo-field~\cite{barnett_gyromagnetic_1935,heims_theory_1962} in the rotating frame that nevertheless has physically significant consequences, such as the Barnett effect~\cite{barnett_magnetization_1915}. Detecting these pseudo-fields is experimentally challenging, as a rotating-frame sensor is required. Previous work has realised classical rotating-frame detectors~\cite{chudo_observation_2014}. Here we use quantum sensors, nitrogen-vacancy (NV) centres, in a rapidly rotating diamond to detect pseudo-fields in the rotating frame. While conventional magnetic fields induce precession at a rate proportional to the gyromagnetic ratio, rotation shifts the precession of all spins equally, and thus primarily affect nearby $^{13}$C nuclear spins.
We are thus able to explore these effects via quantum sensing in a rapidly rotating frame, and define a new approach to quantum control using rotationally-induced nuclear spin-selective magnetic fields. This work provides an integral step towards realising precision rotation sensing and quantum spin gyroscopes.}}
\end{abstract}

\maketitle

A spin measured by an observer in a rotating frame appears to precess faster or slower depending on $\boldsymbol{\Omega}$, the rotational angular frequency of the frame. This can be thought of as arising from an effective magnetic field~\cite{barnett_gyromagnetic_1935,heims_theory_1962} $\mathbf{B}_{\Omega} = \boldsymbol{\Omega}/\gamma$ in the rotating frame, with $\gamma$ the spin gyromagnetic ratio. Despite being referred to as `fictitious' fields, rotationally-induced magnetic pseudo-fields have measurable effects, in the same way that spin-state-dependent vector light shifts~\cite{happer_effective_1967} and artificial gauge fields~\cite{lin_synthetic_2009} have real effects. In the Barnett effect~\cite{barnett_magnetization_1915}, for example, the effective magnetic field generated by physically rotating an initially unmagnetised rod of iron leads to polarisation of the constituent electron spins along the rotation axis, and magnetisation of the iron sample. In this work, we explore for the first time quantum sensing of pseudo-fields in the physically rotating frame, using solid-state qubits that detect rotational pseudo-fields and simultaneously are uniquely suited to exploring quantum control with rotation.     
 
Exploring rotational pseudo-fields imposes considerable experimental challenges, as the sensor must be in the rotating frame~\cite{chudo_observation_2014, lendinez_rotational_2010}. 
Nuclear spin gyroscopes operate on a similar principal, where it is the sensing apparatus that executes rotations about a gas of nuclear spins~\cite{donley_nuclear_2010}. 
Magic-angle spinning~\cite{andrew_magic_1981} nuclear magnetic resonance (NMR) experiments routinely study rapid rotations of more than $10\,$kHz in nuclear spin systems, and recent work has used a pick-up coil rotating with the sample to measure the rotational pseudo-field~\cite{chudo_observation_2014,harii_line_2015,chudo_rotational_2015}. 
However, NMR-based experiments require a strong polarising magnetic field (much larger than the pseudo-fields) to obtain a signal, limiting these experiments to detection of small perturbations due to rotation.
Both nuclear spin gyroscopes and pickup coils are operated essentially classically, with limited scope to fully study the effects of rotational pseudo-fields on quantum systems. 

Solid-state spin systems, such as the nitrogen-vacancy (NV) centre in diamond~\cite{jelezko_single_2006,doherty_nitrogen-vacancy_2013,schirhagl_nitrogen-vacancy_2014} have attracted considerable attention as robust quantum sensors, and have innate advantages for the study of rotational pseudo-fields. NV centres are intrinsic to the rotating sample, with an electron spin that is easily controlled and measured with microwave and optical fields, and are also amenable to quantum measurement and control protocols \cite{prawer_quantum_2014}. Nuclear spins (such as spin-1/2 $^{13}$C) located within a few lattice sites of the NV spin precess at kHz rates in laboratory fields of several gauss, rotation rates achievable with modern electric motors and magic-angle spinners. The electron spin coherence time $T_2\sim 0.1-1\,$ms of the NV centre~\cite{stanwix_coherence_2010, balasubramanian_ultralong_2009} can be made comparable to the nuclear spin precession period at low fields, allowing a substantial time window for quantum sensing during rotation. The NV electron spin directly measures the nuclear spin magnetic dipole field, so we do not need strong conventional fields to polarise nuclear spins to attain a measurement signal.

A conventional magnetic field induces precession at a rate proportional to the spin gyromagnetic ratio, but rotational pseudo-fields shift the precession of all spins equally, the effective field being inversely proportional to the gyromagnetic ratio. In addition to quantum detection, a rotating NV$-$nuclear spin system allows our experiments to investigate a regime denied to previous studies, where rotational pseudo-fields are large enough to cancel a conventional magnetic field for the nuclear spins. The NV electron spins are essentially unaffected by rotation, and retain a significant Zeeman splitting from the conventional field. The NV remains an independent, controllable and incisive probe of the zero-field nuclear spin dynamics.

While the NV has been used as a quantum sensor in a variety of noisy, real-world conditions, such as within biological cells~\cite{mcguinness_quantum_2011, kucsko_nanometre-scale_2013}, it is not immediately apparent that its abundant sensing advantages can be accessed when executing rapid rotation. A physically rotating NV centre is predicted to accumulate a geometric phase~\cite{maclaurin_measurable_2012}, which forms the basis of a proposed nanoscale gyroscope~\cite{ledbetter_gyroscopes_2012}. Theoretical work has also proposed a nuclear-spin gyroscope using the intrinsic nitrogen nuclear spin of the NV centre to sense rotationally-shifted precession~\cite{ajoy_stable_2012}. To date, experiments on NVs in moving diamonds have considered the quasi-static case, where standard experimental protocols can be applied~\cite{mcguinness_quantum_2011}. In our experiments, we employ NV centres as quantum magnetometers in a frame rotating with a period comparable to the spin coherence time, and establish the means of extracting quantum information from rapidly rotating qubits.

\begin{figure*}[t!]
	\centering
		\includegraphics[width = \textwidth]{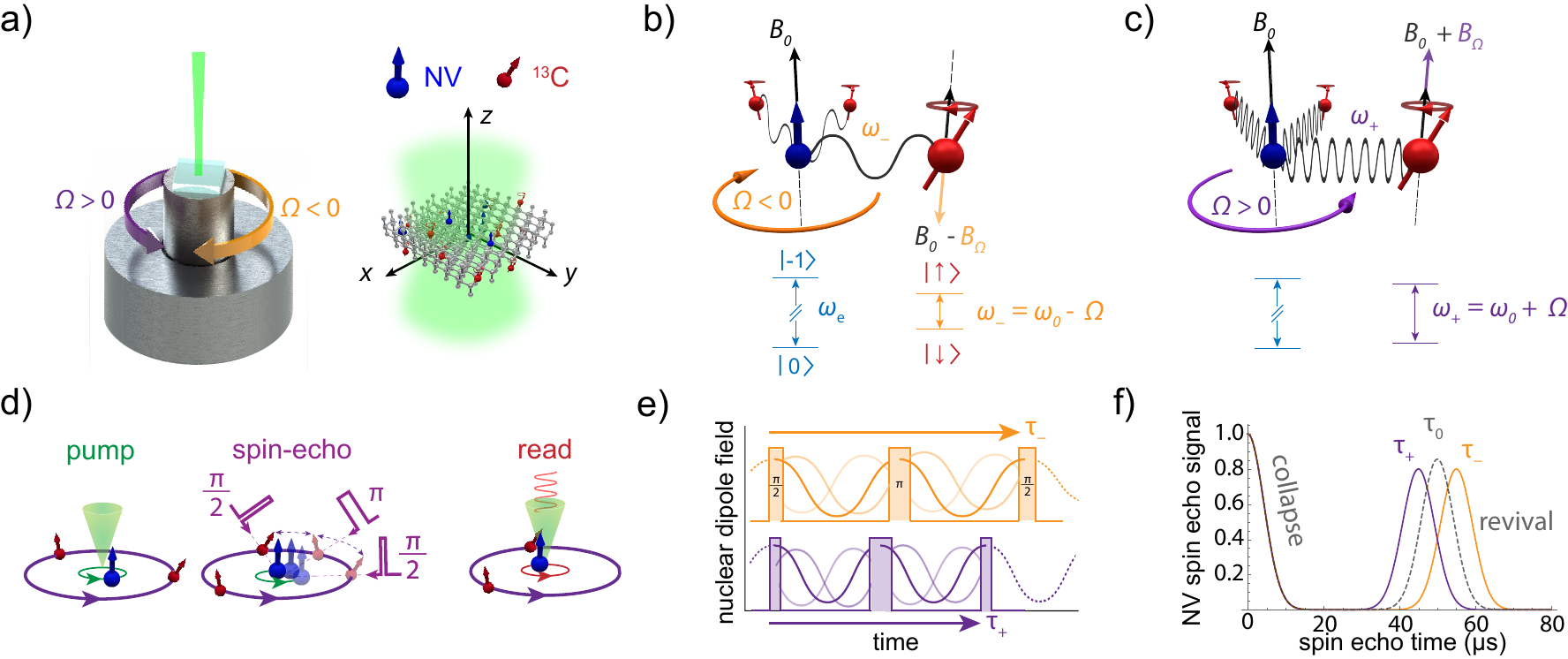}
	\caption{{\bf Experimental setup and measurement scheme.} {\bf{a}}, A diamond containing an ensemble of NV centres and natural abundance $^{13}$C is mounted on an electric motor spindle such that one of the NV orientation classes is parallel to the rotation axis $\uvect{z}$. NV electron spins interact with nearby $^{13}$C nuclei, experiencing a time-varying magnetic field from the precessing nuclear magnetic dipoles. Depending on the relative direction of rotation and Larmor precession, the rotationally-induced pseudo-field $\vect{B}_{\Omega}$ decreases ({\bf{b}}) or increases ({\bf{c}}) the frequency  of the $^{13}$C dipole field as seen by the NV sensor. The strength of the pseudo-field is inversely proportional to the gyromagnetic ratio: the $^{13}$C nuclei experience a large $\vect{B}_{\Omega}$ from rotation, whereas the much smaller electron field (not shown) due to $\gamma_e \gg\gamma_{13\text{C}}$ means the electron spin is essentially unperturbed. {\bf d}, Schematic of the experimental sequence, the NV centres are optically prepared before microwave pulses manipulate the NV spins in a spin-echo sequence. A final laser pulse reads the NV spin state. {\bf e}, The $^{13}$C dipole field is measured in a spin-echo experiment. {\bf{f}}, The spin-echo signal at first collapses, then revives when the measurement time is equal to two periods of the nuclear spin precession. The pseudo-field in a given rotation configuration is detected by by determining the time $\tau_+, \tau_-$ the echo revival occurs. In the second half of this work, we observe the effects of rotation on the initial collapse of the echo signal.}
	\label{fig:master}
\end{figure*}

An outline of our experiment is depicted in Figure \ref{fig:master}. For simplicity, we show the case for single NV sensors, but the physics is equally valid for the ensembles of NVs used in our experiments: multiple identical NV sensors greatly enhance the measurement signal relative to noise and considerably simplify the experimental procedure. A synthetic diamond with an ensemble density of NVs and a $1.1\%$ natural abundance of $^{13}$C is mounted to a high-speed electric motor, with one of the four orientation classes of NVs approximately parallel to the rotation axis (denoted as $\uvect{z}$). With a magnetic field $\vect{B}_0$ also parallel to $\uvect{z}$, an effective two-level system is formed from the $m_S = 0$ and $m_S = -1$ states of the NV ground state. The experimental procedure, summarised in Figure \ref{fig:master}{\bf d}, consists of standard optical preparation, microwave state manipulation and analysis of time-dependent photoluminesence emitted by the ensemble to determine the relative phase between the states of the NV two-level system accumulated in the spin-echo sequence. We employ several modifications to accommodate the rotation of the diamond (see Methods).

The first part of this work concerns detection of rotational pseudo-fields. Spin-1/2 $^{13}$C nuclei in the diamond lattice (precessing at ${\omega_{13\text{C}} = 2\pi f_{13\text{C}} = \gamma_{13\text{C}} B }$ in a magnetic field of strength $B  = |\vect{B}|$, $\gamma_{13\text{C}}/2\pi = 1.0715\,\text{kHz/G}$) generate a time-varying magnetic field at the NV that is detected in a spin-echo experiment~\cite{childress_coherent_2006}. With the NV in the $m_S = 0$ state and the magnetic field parallel to the NV axis, the nuclear spins precess at $f_{13\text{C}}$. When the NV is in the $m_S = -1$ state, its dipole field interacts with the $^{13}$C spin, rapidly modulating the spin-echo signal. The many different configurations of NV--$^{13}$C pairwise interactions, when averaged over the ensemble, gives a spread of oscillation frequencies that in turn beat against each other. The spin echo signal collapses with a magnetic field-strength dependent characteristic time\cite{childress_coherent_2006} $\tau_C (B)$, $S(t)\propto \exp\left(-(t/\tau_C)^4\right)$. When the spin-echo measurement time $\tau$ equals $2/f_{13\text{C}}$, the total phase accumulated is zero, and the spin-echo signal revives. Any change in the nuclear spin precession frequency changes the time at which the echo signal revives. We measure this revival time to infer the overall $^{13}$C precession frequency $f_{13\text{C}}$ and then quantify the rotationally-induced shift (Figure \ref{fig:master}{\bf d}).

The gyromagnetic ratio of the $^{13}$C nucleus is positive: in the presence of a magnetic field oriented along $+\uvect{z}$, the $^{13}$C dipole moment precesses in a negative direction (clockwise, looking from above along $-\uvect{z}$, as shown in Figure \ref{fig:master}{\bf{b},\bf{c}}). The $^{13}$C spin state populations are thermally distributed in our experiments, but it is the relative direction of spin precession and physical rotation that is significant. The precession direction of the $^{13}$C dipole moment is the same, regardless of spin state. Inverting the magnetic field direction changes the precession direction of the $^{13}$C dipole relative to the imposed rotation $\Omega = 2\pi f_\text{rot}$. There are thus four possible configurations we investigate: $\vect{B}_0 = \pm B_0\,\uvect{z}$ and anticlockwise ($f_\text{rot}>0$) or clockwise ($f_\text{rot}<0$) rotations.

\begin{figure}[t!]
	\centering
		\includegraphics{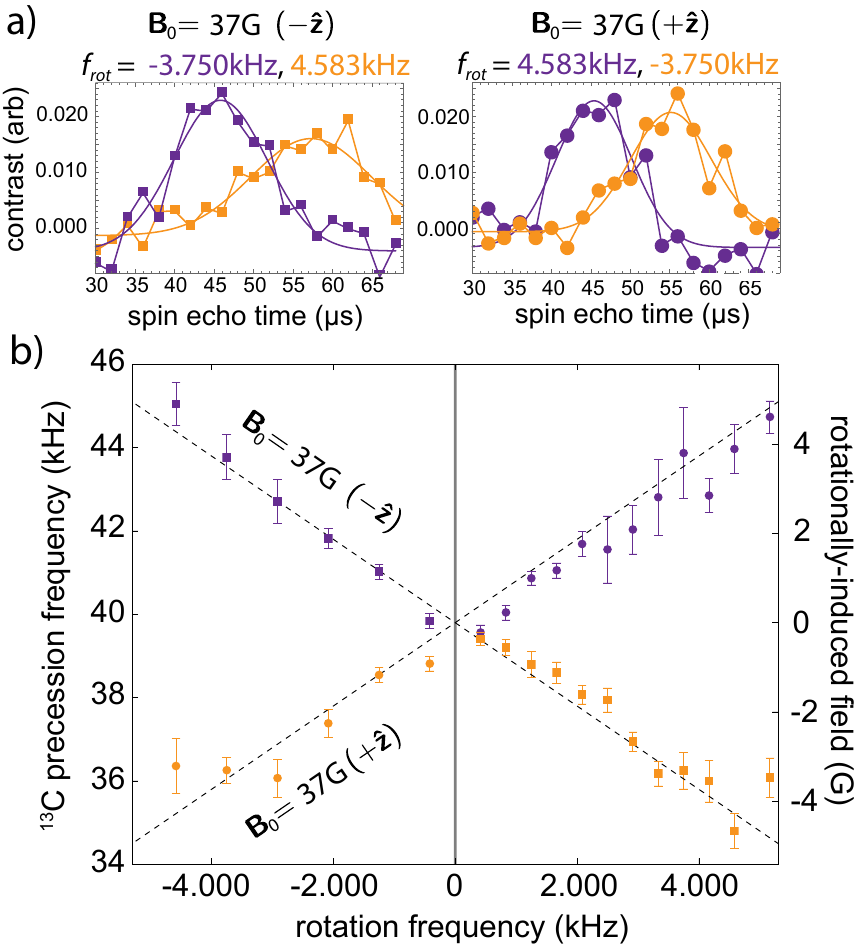}
	\caption{{\bf Measurement of rotationally-induced magnetic pseudo-fields}. Rotational pseudo-fields increase or decrease the precession frequency of the $^{13}$C nuclei, and thus change the time when the spin-echo signal revives. {\bf a} Example echo revival signals and Gaussian fits at different rotation speeds and magnetic field directions. {\bf b} The $^{13}$C precession frequency is determined from revival times extracted from Gaussian fits to the echo signal and plotted as a function of rotation frequency for different field orientations. The dashed black lines denote $f_{13\text{C}} = f_0 \pm f_\text{rot}$, a $^{13}$C precession frequency perturbed only by the rotation frequency of the diamond. Error bars derived from standard error of the fitted Gaussian centroids.}
	\label{fig:rotocross}
\end{figure}

The total field $\vect{B} = \vect{B}_0+\vect{B}_\Omega$ experienced by the $^{13}$C spins is measured by determining the time when the spin-echo signal revives. The echo signal around the revival typically appears Gaussian in time, and for each rotation speed and magnetic field configuration we fit a Gaussian to the echo signal and extract the revival time. Example revivals are shown in Figure \ref{fig:rotocross}{\bf a}. The extracted $^{13}$C precession frequency for an applied field of $\vect{B}_0=\pm B_0\uvect{z}$ ($B_0=37\,$G, $f_0 = 40\,$kHz) is shown in Figure \ref{fig:rotocross}{\bf b} as a function of rotation speed for the four possible configurations of rotation and magnetic field. The results closely match the expected linear shift $f_{13\text{C}} = f_0\pm f_\text{rot}$ characteristic of rotational pseudo-fields.   

\begin{figure*}
	\centering
		\includegraphics[width = \textwidth]{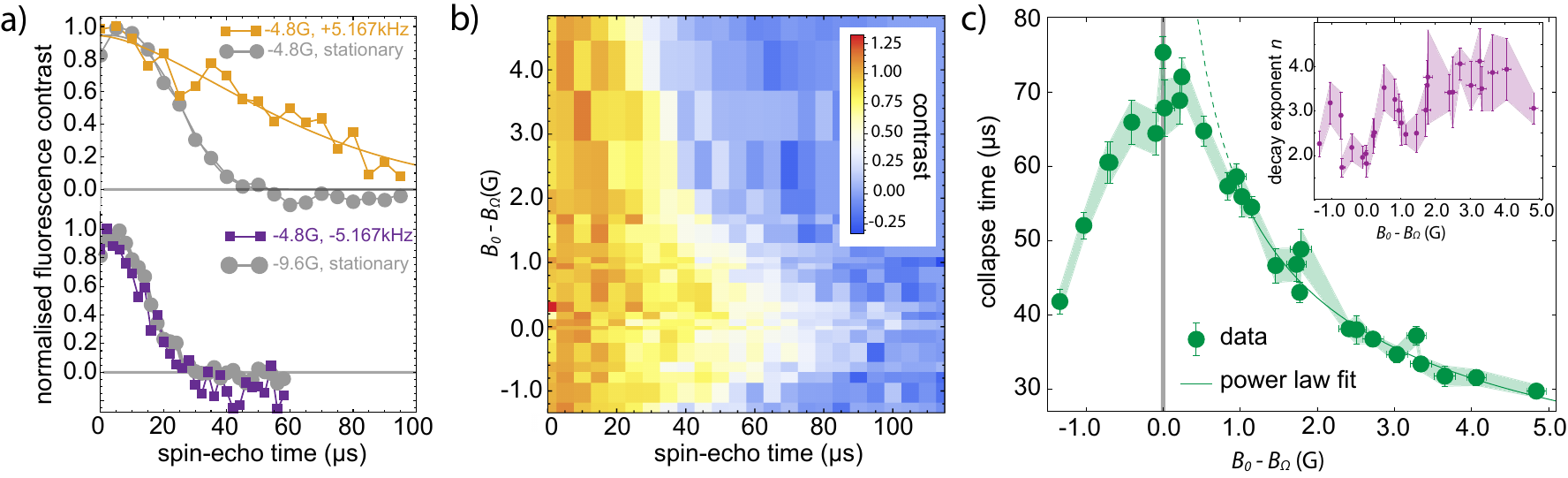}
	\caption{{\bf Quantum control of nuclear spins with rotational pseudo-fields}. {\bf a} The initial collapse of the spin-echo signal depends on the strength of the total field: for a $4.8\,$G bias field along $-\uvect{z}$ (echo signal shown in grey), rotation at $5.167\,$kHz cancels the applied magnetic field (orange), leading to slower collapse of the echo signal, whereas rotation at $-5.167\,$kHz results in a total field of $B \approx 9.6\,$G (purple), closely matching that observed with a stationary diamond and the same total field strength (grey). {\bf b} Spin-echo signal as a function of $B_0-B_\Omega$. {\bf c} We fit Eq. \ref{eq:decay} to the echo signal and plot the extracted collapse time $\tau_C$ and decay exponent $n$ (inset). The fitted $\tau_C$ follows the predicted power law behaviour~\cite{zhao_decoherence_2012} for $B>1\,$G before saturating to a value of $\tau_C \approx 70\,\upmu$s near zero field. For fields above 2\,G, the decay exponent is steady around $n=4$, and at zero field is approximately $n = 2$ but exhibits scatter due to the limited number of points constraining the fitted value of $n$. Error bars are the parameter standard errors derived from least square fits to the data in panel {\bf b}.}
	\label{fig:rolloff}
\end{figure*}

We then considered the case where the rotational pseudo-field is comparable to the external magnetic field, allowing us to cancel the conventional magnetic field for the nuclear spins in the rotating frame and thus control the NV electron spin coherence. We studied the initial collapse of the spin-echo signal, where the field-dependent collapse time $\tau_C(B)$ is indicative of the dominant spin bath interactions~\cite{zhao_decoherence_2012}: for $B>1\,$G, the NV-$^{13}$C hyperfine interaction dominates and the collapse time is expected to increase with lower magnetic field strengths. Below $1\,$G, nuclear spin flips mediated by the internuclear dipole-dipole interaction dominate, with $\tau_C$ saturating to some maximum value limited by the uncorrelated magnetic noise in the spin bath.

In Figure \ref{fig:rolloff}{\bf a } we show the initial collapse of the echo signal for an applied $B_0 = 4.8\,$G field along $-\uvect{z}$. With the diamond rotated at $f_\text{rot} = 5.167$\,kHz, the induced $\vect{B}_\Omega$ is along $+\uvect{z}$ and cancels the magnetic field (top). When rotated at $f_\text{rot} = -5.167$\,kHz, $\vect{B}_\Omega$ is parallel to $\vect{B}_0$ resulting in $B = 9.6\,$G, the measured spin-echo signal closely matches the echo signal observed for a stationary diamond and an applied $9.6\,$G field. We then observed how the echo signal changes for a range of total fields, from $B_0-B_\Omega = -1.34(4)\,$G to $4.84(1)\,$G (Figure \ref{fig:rolloff}{\bf b}). We modelled the normalised spin-echo signal as~\cite{stanwix_coherence_2010} 
\begin{equation}
S(t) = \exp\left(-(t/\tau_C)^n\right),
\label{eq:decay}
\end{equation}
with free collapse time $\tau_C$ and decay exponent $n$. The $\tau_C$ extracted from fits to the data is shown in Figure \ref{fig:rolloff}{\bf c}. The collapse time is well described by the power law function $\tau_C\propto B^{-k}$ with $k = 0.42(2)$ for $B < 1$\,G before saturating at $B = 0$ to around $70\,\upmu$s. The value of $k$ in the power law model is theoretically predicted~\cite{zhao_decoherence_2012}  to be $0.5$ (for a single NV in a nuclear spin bath), but depends on the specifics of the bath environment. The collapse time is relatively independent of the decay exponent. The value of $n$ is also expected to change with field strength, from $n= 4$ to $n= 2$ as the field is reduced from moderate strength to near zero~\cite{hall_analytic_2014}. While we did observe this general trend, improved statistics would be needed for more detailed investigation of the behaviour of the decay exponent.    

The NV electron spin coherence is intrinsically linked to the dynamics of the surrounding bath of nuclear spins. We have shown in Figure \ref{fig:rolloff} that pseudo-fields allow us to control the spin bath independently of the NV electron spin, which retains its sensing utility even when the total field experienced by the nuclear spins is zero. This nuclear-spin selectivity stems from a unique property of rotational pseudo-fields: $B_{\Omega, e} = \Omega/\gamma_e$, experienced by the NV (with gyromagnetic factor $\gamma_e/2\pi = 2.8\,\text{MHz\,G}^{-1})$ is a factor of $\gamma_{13\text{C}}/\gamma_e$ less than $B_{\Omega, \text{13C}}$, that experienced by the $^{13}$C spins. For the maximum rotation frequency used in this work, $5.5\,$kHz, $B_{\Omega, e} =2\,$mG compared to $B_{\Omega, \text{13C}} =5.13\,$G for the $^{13}$C spins, and for all rotation speeds considered we did not observe any additional Zeeman splitting of the NV energy levels. 

The nuclear-spin selectivity of pseudo-fields invites application in other schemes to selectively manipulate nuclear spins in diamond. For instance, much higher rotation speeds (111\,kHz has been demonstrated in NMR magic angle spinning experiments~\cite{andreas_structure_2016}) could create $^{13}$C-specific fields of more than 100\,G while still minimally perturbing the NV electron spin. Large pseudo-fields potentially offer alternatives to existing decoupling schemes, controlling different nuclear spin bath elements independently of the NV electron spin and enriching zero-field nanoscale NMR experiments with NV centres~\cite{staudacher_nuclear_2013,sushkov_magnetic_2014,lovchinsky_nuclear_2016} and schemes of quantum information processing between coupled electron and nuclear spins~\cite{dutt_quantum_2007}. 

In addition to using quantum sensors to demonstrate the profound connection between magnetism and physical rotation, our findings advance quantum sensing and measurement into the physically rotating frame, and form a significant step towards NV-based rotation sensing and gyroscopic applications. Our results also establish a unique, highly selective method of controlling the nuclear spin bath surrounding an ensemble of electron spin qubits, and we anticipate interesting new directions for employing rotation as a quantum control. 

\section*{Methods}
{\bf State preparation and readout.} The diamond used in this work contains an ensemble density of NV centers equally distributed between the four possible orientation classes, one of which is normal to the diamond surface. The $\uvect{z}-$oriented magnetic bias field is aligned parallel to the rotation axis, so that only the NV centers normal to the surface of the diamond are resonantly addressed with microwaves. The preparation laser is aligned close to the center of rotation to maximise optical pumping and readout efficiency. For rotational speeds above 1.667\,kHz the NVs are illuminated with green light for one rotational period, optically preparing a ring of NV centers. At lower speeds the NVs are prepared using a $3\,\upmu$s laser pulse; at these speeds drift of the rotational center during experiments is less severe. We verified the efficacy of optical preparation for both pumping schemes. Since the NV axis is parallel to the rotation axis, each microwave pulse then addresses the entire ring of optically prepared NVs, and precise synchronisation to the motor rotation is not necessary.

{\bf Magnetic field alignment.} The NV orientation class we probe is not precisely aligned with the rotation axis. When the magnetic field is also tilted from the rotation axis, the Zeeman shift of the $m_S=\pm1$ states changes during rotation. This leads to an effective AC field as experienced by the NVs of magnitude $B_\text{eff} = B_0 \sin\theta_B\sin\theta_\text{NV}$, with $B_0$ the total field strength and $\theta_B$ and $\theta_\text{NV}$ the misalignment of the magnetic field and NV axis from the rotation axis, respectively. A spin-echo measurement is sensitive to these fields, which in the case of large misalignments suppress the $^{13}$C revival or result in faster initial collapse, since the experimental pulse sequence is not synchronous with the rotation of the diamond. We use the amplitude of the spin-echo signal to diagnose the elimination of magnetic field components transverse to the rotation axis and minimise $\theta_B$. 

\section*{Acknowledgements}
We acknowledge valuable discussions with L. P. McGuinness, J.-P. Tetienne, D. A. Simpson and A. D. Stacey for provision of the diamond sample. This work was supported by the Australian Research Council Discovery Scheme (DP150101704). 
\section*{Author Contributions}
Experiments were performed and analysed by A.A.W and E.L; A.A.W., Y.Y.F. and R.E.S. designed and constructed the experimental apparatus. L.C.L.H., V.S.P. and A.M.M. conducted the theoretical investigation. A.A.W., L.C.L.H., R.E.S. and A.M.M. conceived the experiment and wrote the manuscript with contributions from all authors.
\section*{Additional information}
The authors declare no competing financial interests.
Correspondence and requests for materials should be addressed to A.~M.~M.

\end{document}